\documentclass[preprint,aps,nofootinbib]{revtex4-1}

\usepackage{dcolumn}
\usepackage{bm}
\usepackage{epsfig}
\usepackage{graphicx,epstopdf}
\usepackage{float}
\usepackage{paralist}

\usepackage{amsmath}
\usepackage{amsfonts}
\usepackage{amssymb}
\usepackage{adjustbox}
\usepackage{color}
\usepackage{xcolor}
\usepackage{adjustbox}
\usepackage{subcaption}
\usepackage[toc,page]{appendix}
\usepackage[utf8]{inputenc}
\usepackage{slashed}
\usepackage{indentfirst}
\usepackage[justification=justified,singlelinecheck=false]{caption}
\usepackage{subcaption}
\usepackage{hyperref} 
\usepackage{natbib} 
\usepackage{ulem}
\def\tt{\tilde{t}}
\def\lsp{\tilde{\chi}^0_1}
\def\mtt{m_{\tilde{t}}}
\def\mchar{m_{\tilde{\chi}_1^{\pm}}}
\def\mneutwo{m_{\tilde{\chi}_2^{0}}}
\def\msneu{m_{\tilde{\nu}}}
\def\mslep{m_{\tilde{\ell}}}
\def\mlsp{m_{\tilde{\chi}_1^{0}}}
\def\Rmb{\bar{R}_M}
\def\Rmbmax{\bar{R}_{\text{max}}}
\def\Rmbmin{\bar{R}_{\text{min}}}
\def\ifb{\text{fb}^{-1}}

\begin{document}

\title{
Compressed Stop Searches with Two Leptons and Two $b$-jets
}
\author{Hsin-Chia Cheng,$^{a,b}$ Christina Gao,$^a$ Lingfeng Li$^{a,c}$}
\email[Email: ]{cheng@physics.ucdavis.edu}\email{cygao@ucdavis.edu}\email{llfli@ucdavis.edu}
\affiliation{$^a$Department of Physics, University of California Davis, Davis, California 95616, USA\\
$^b$School of Natural Sciences, Institute for Advanced Study, Princeton, New Jersey 08540, USA\\
$^c$Jockey Club Institute for Advanced Study, Hong Kong
University of Science and Technology, Hong Kong}

\begin{abstract}
In top squark (stop) searches with a compressed spectrum, it is very helpful to consider the stop production recoiling against a hard jet from the initial state radiation to obtain a significant amount of missing transverse energy. In particular, the kinematic variable $R_M$ which measures the ratio of the lightest neutralino mass and the stop mass proved to be crucial in separating the signals from the backgrounds in both the all-hadronic decay and the semileptonic decay of the stops. Here we generalize the search method to the dileptonic stop decays. In this case, due to the two missing neutrinos, there are not enough kinematic constraint equations to solve for the $R_M$ variable exactly, but only render an allowed interval consistent with the event. However, we show that the minimum and the maximum values of this allowed interval still provide useful variables in discriminating signals from the backgrounds. Although in the traditional stop decay to a top quark and the lightest neutralino, the dileptonic mode is not as competitive due to its small branching ratio, it becomes the main search mode if the stops decay through the charginos and sleptons with a compressed spectrum. We show that with the new variables, the dileptonic search of the stop can cover regions of the parameter space which have not been constrained before.   
\end{abstract}
\maketitle

\section{Introduction}

Since the discovery of a 125 GeV Higgs boson in 2012~\cite{Aad:2012tfa,Chatrchyan:2012xdj}, the Large Hadron Collider (LHC) has not discovered any other new elementary particle or observed major deviations from the Standard Model (SM). It leaves the naturalness problem of the SM still a mystery, as the large quadratic contribution to the Higgs mass-squared parameter from the top quark loop would destabilize the electroweak (EW) scale if it is not canceled.  A major theoretical endeavor to address the naturalness problem is to extend the SM by supersymmetry (SUSY), so that the Higgs mass is protected by this additional symmetry from the quadratic divergence. Under SUSY, every SM fermion (boson) has its bosonic (fermionic) partner. The superpartners must receive large enough masses from SUSY breaking effects so that they have escaped the experimental detection so far. 
Among them, the superpartners of the top quark (top squarks or stops) are most relevant for the naturalness problem due to the large top loop contribution to the Higgs. They should not be far above the weak scale in order to cut off the quadratic top loop contibution.
Another benefit of SUSY is that it provides natural candidates for dark matter in the universe if the $R$-parity is conserved. The lightest supersymmetric particle (LSP) is stable and can be a weakly interacting massive particle (WIMP) dark matter if it is not charged under $U(1)_{\rm EM}$ or $SU(3)_C$, e.g., the lightest neutralino which is a linear combination of the superpartners of the EW gauge bosons and  the Higgs boson. At colliders, superparticles are pair-produced and decay down to the LSPs which escape the detectors, leaving missing energy signals as one of the signatures of SUSY. 

There have been extensive searches for the stops at the LHC in various channels. With the new Run 2 results, CMS and ATLAS~\cite{ATLAS:2017kyf,ATLAS:2017tmd,ATLAS:2017msx,
Sirunyan:2017cwe,Sirunyan:2017cwe,CMS:2017qjo,CMS:2017arv,
CMS:2017vbf} have pushed the lower limit of the stop mass to  $\sim 1$ TeV, assuming that the stop decays to the LSP $\tilde{\chi}_1^0$ and the top, and that $m_{\tilde{\chi}_1^0}\lesssim 200$~GeV. The limit becomes weaker for a smaller mass difference between the stop and the $\tilde{\chi}_1^0$, or if the stop decays differently in the case that the LSP is the sneutrino~\cite{Chala:2017jgg}. For the difficult cases of very compressed spectra, $\mtt \lesssim m_t+\mlsp$, various search channels and kinematic variables have been proposed~\cite{Hikasa:1987db,Muhlleitner:2011ww,Boehm:1999tr,Das:2001kd,
Drees:2012dd,Bai:2013ema,Carena:2008mj,Hagiwara:2013tva,
An:2015uwa,Macaluso:2015wja,Cheng:2016mcw,Jackson:2016mfb,Konar:2016ata} and new techniques have been adopted in recent experimental searches~\cite{ATLAS:2017kyf,ATLAS:2017tmd,ATLAS:2017msx,
Sirunyan:2017cwe,Sirunyan:2017cwe,CMS:2017arv,
CMS:2017vbf}.

The search in the compressed region has been challenging because its signature is hard to distinguish from the SM $t\bar{t}$ production at the LHC. 
For $m_{\tilde{t}_1} = m_{t}+m_{\tilde{\chi}_1^0}$ which is called the top corridor, the LSP and top are almost static in the rest frame of the stop decay. Therefore, in the lab frame, the top and the LSP would be collinear and that
\begin{equation}\label{mass}
\frac{p_{\tilde{\chi}_1^0}}{p_{\tilde{t}_1}} \approx \frac{m_{\tilde{\chi}_1^0}}{m_{\tilde{t}_1}}.
\end{equation} 
In the stop pair production, the two LSPs tend to travel back to back, resulting in a cancellation of their transverse momenta, thus leaving little trace for $\tilde{\chi}_1^0$s.
A way to separate the signals from the backgrounds is to consider the stop pair production together with a hard jet from the initial state radiation (ISR)~\cite{Carena:2008mj,Hagiwara:2013tva,An:2015uwa,Macaluso:2015wja,Cheng:2016mcw,Jackson:2016mfb}.  From the conservation of momentum, 
\begin{equation}\label{pt}
p_{T(J_{\rm ISR})} \approx -\sum_{i=1}^2 p_{T\tilde{t}_{1,i}}
\end{equation}
both LSPs tend to be emitted antiparallel to the ISR jet, resulting in a significant amount of missing transverse momentum ($\slashed{p}_T$).
By studying the fully hadronic decays of such events, it was pointed out that the ratio between $\slashed{p}_T$ and $p_{T(J_{\text{ISR}})}$, defined as $R_M$, can be a useful kinematic variable to differentiate between the stop and top decays~\cite{An:2015uwa,Macaluso:2015wja}. Since the missing momentum in stop decays is mainly due to the presence of LSPs, together with equations (\ref{mass}) and (\ref{pt}), the stop pair production tends to have  
\begin{equation}\label{RM}
R_M\equiv\frac{\slashed{p}_T}{p_{T(J_{\text{ISR}})}}\approx \frac{m_{\tilde{\chi}_1^0}}{m_{\tilde{t}_1}}\,  ,
\end{equation} 
which is expected between zero and one. On the other hand, the $\slashed{p}_T$ for the hadronic decays of $t\bar{t}$ is due to the detector smearing effect, so $R_M$ is expected close to zero for the background. Using this variable, with the help of the recursive jigsaw method~\cite{Jackson:2016mfb} to separate the ISR from the stop system, ATLAS has excluded the stop mass between 235--590 GeV along the top corridor ($m_{\tilde{t}_1} \approx m_{t}+m_{\tilde{\chi}_1^0}$) with 36.1 fb$^{-1}$ of integrated luminosity~\cite{ATLAS:2017kyf}.

For the semileptonic and dileptonic decays of the stops, $R_M$ becomes less informative if the neutrinos' contribution to $\slashed{p}_T$ cannot be easily separated from that of neutralinos. In general for the same $p_{T(J_{\text{ISR}})}$, the signal events are still expected to have larger missing transverse energies (MET) than those of the backgrounds. This can be used to constrain the compressed stop, though the current bounds are somewhat weaker than the all-hadronic channel~\cite{ATLAS:2017tmd,ATLAS:2017msx,CMS:2017arv}. In a previous paper~\cite{Cheng:2016mcw} we showed that for stop semileptonic decays, the neutrino momentum can be solved (up to a two-fold ambiguity) by exploiting the kinematic features of the compressed stop spectrum in the top corridor. Therefore the neutrino momentum can be subtracted from $\slashed{p}_T$ and a modified version of $R_M$ can be defined:
\begin{equation}\label{RMbar}
\bar{R}_M\equiv\frac{\slashed{p}_{T(\chi)}}{p_{T(J_{\text{ISR}})}}\approx\frac{\slashed{p}^{\parallel}_{T}-p^{\parallel}_{T\nu}}{p_{T(J_{\text{ISR}})}}\approx \frac{m_{\tilde{\chi}_1^0}}{m_{\tilde{t}_1}},
\end{equation} 
where $p^{\parallel}_{T\nu}$ is the neutrino's contribution to the missing momentum along the ISR direction. The $\bar{R}_M$ variable can be similarly used in the stop search along the top corridor for signal events with one lepton.

\begin{figure}[t]
\captionsetup{singlelinecheck = false, format= hang, justification=raggedright, font=footnotesize, labelsep=space}

\includegraphics[scale=0.8]{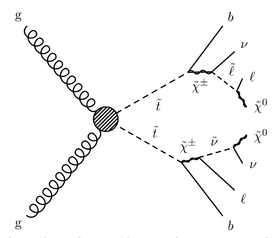}
\caption{Possible stop/chargino decay chains for the stop-chargino-slepton scenario. Both stops decays to $b+\tilde{\chi}^{\pm}$. There are two possible chargino decays through sleptons. The top leg shows the chargino decaying to a charged slepton and an SM neutrino, then the slepton further decays to a SM charged lepton and the LSP. The decay of the chargino in the bottom leg gives a SM charged lepton and a sneutrino, the latter then decays invisibly to a neutrino and the LSP. For a Wino-like chargino and a nearly degenerate slepton-sneutrino pair, the branching ratios of these decay are approximately equal.}
\label{fig:feyndiag}
\end{figure}

The dileptonic decays of the stops have been considered as a useful way for stop searches since early times~\cite{Baer:1988vk,Baer:1991cb,Baer:1994xr}. However, in the top corridor
the situation is tougher. The branching fraction is small if the stops decay through the tops. Furthermore, there are more unknowns than the number of kinematic constraints available from the decay chain, hence the two neutrinos' contribution to $\slashed{p}_T$ cannot be solved exactly and be completely removed from the total $\slashed{p}_T$. One would think that the final states with $b\bar{b}\ell^+\ell^- +\slashed{p}_T$ may not be a competitive channel for the compressed stop search. However, with a more complicated spectrum, there could be other decay chains of stops which end up with the same dileptonic final states and they can be even dominant in certain cases. For example, this can occur if there are also charginos and sleptons in the spectrum between the stop and the lightest neutralino as shown in Fig.~\ref{fig:feyndiag}. The all hadronic and semileptonic searches may become ineffective if the dileptonic decays become dominant. Therefore, it is still worthwhile to consider the dileptonic search channel for the compressed stop.
Another potential merit of the dileptonic channel is that 
the only significant background to the $b\bar{b}\ell^+\ell^- +\slashed{p}_T$ final states is the $t\bar{t}$ dileptonic decays, so one only needs to focus on suppressing this background. In contrast, the all hadronic and the semileptonic signals also suffer from $t\bar{t}$ decays with an extra lepton which is not identified, and they are often the major backgrounds in the $R_M$ variable. 

Despite the fact that the two neutrinos' momenta can not be solved exactly for the dileptonic events from the $t\bar{t}$ or the stop pair in the top corridor with ISR, the kinematic constraints still strongly limit the ranges of their possible values, which translate to an allowed range of $\bar{R}_M$ values for each $2\ell$2$b+\slashed{p}_T+$jet(s) final state event. For the $t\bar{t}$ background, $\bar{R}_M$ is expected to be close to zero and contain the point zero. For the stop events, due to the additional $\tilde{\chi}_1^0$'s contribution to $\slashed{p}_T$, it is expected to shift to larger values. Even for stop events decaying through sleptons as in Fig.~\ref{fig:feyndiag}, which do not have the correct  kinematics for the kinematic constraint equations of $t\bar{t}$, one can still obtain the corresponding $\bar{R}_M$ ranges anyway, and there is no reason that they should be close to zero as the $t\bar{t}$ background. As a result, the allowed $\bar{R}_M$ range, characterized by its minimum and maximum values $\bar{R}_{\text{min}}$, $\bar{R}_{\text{max}}$, may be used to distinguish the signals from the backgrounds of the $2\ell$2$b+\slashed{p}_T+$jet(s) final state, irrespective of the stop decay topologies. 

The goal of this work is to study the usefulness of the $\bar{R}_M$ variables in the dileptonic decay channels of the compressed stop search and their search reaches.
The rest of the paper is organized as follows. 
In section~\ref{sec:kinematics}, we review $\bar{R}_M$ for the semileptonic decays and generalize the concept to dileptonic decays. Even though a unique $\Rmb$ cannot be obtained in the dileptonic case because of insufficient kinematic constraints, the minimum and maximum allowed $\Rmb$ values from the constraints can still provide useful variables to distinguish signals and backgrounds. In section~\ref{Section:Bench}, we perform analyses using the $\bar{R}_{\text{min}}$, $\bar{R}_{\text{max}}$ variables on the chosen benchmarks, for both the stop-slepton decay and the tradition stop dileptonic decay scenarios. Section~\ref{sec:conclusions} contains our conclusions. A detailed description on how we calculate $\bar{R}_{\text{min}}$, $\bar{R}_{\text{max}}$, and a comparison of significances in search analyses with and without using the $\bar{R}_{\text{min}}$, $\bar{R}_{\text{max}}$ variables are collected in appendices.

\section{Kinematics and Variables}
\label{sec:kinematics}

Since the assumptions and tools we employ to analyze the dileptonic stop decays share many similarities to the semileptonic stop decays, we begin by
reviewing the concept of $\bar{R}_M$ in semileptonic signals following Section 2 of Ref.~\cite{Cheng:2016mcw}. 
For the stop pair production with a \emph{hard} ISR jet in the top corridor, $\slashed{p}_T$ due to the neutralinos is approximately antiparallel to the $p_T$ of the ISR, 
because the center of momentum frame of the two neutralinos are the same as the center of momentum frame of the two stops from Eq.~(\ref{mass}).
As a result, the component of $\slashed{p}_T$ perpendicular to the ISR can be attributed to the presence of the neutrino. 
Once the $J_{\text{ISR}}$ is identified, $p_{T\nu}^{\perp}$ is uniquely determined from the experimental measurements. Combining it with the three mass-shell conditions
\begin{equation}\label{mass_topdecay}
\begin{split}
p_\nu^2 &= 0 ,\\
(p_\ell+p_\nu)^2 &= m_{W}^2, \\
(p_\ell+p_\nu+p_b)^2 &= m_t^2,
\end{split}
\end{equation}
and the measured momenta of the lepton and the $b$-jet (assuming that the $b$-jet from the corresponding top decay can be identified), we can solve for the neutrino momentum $p_\nu$ (up to a two-fold ambiguity due to the quadratic mass-shell equations).
After obtaining $p_\nu$, we can subtract its contribution from $\slashed{p}_{T}^{\parallel}$ and get the relation Eq.(\ref{RMbar}):
\begin{equation*}
\bar{R}_M\equiv\frac{\slashed{p}_{T(\tilde{\chi})}}{p_{T(J_{\text{ISR}})}}\approx\frac{\slashed{p}^{\parallel}_{T}-p^{\parallel}_{T\nu}}{p_{T(J_{\text{ISR}})}}\approx \frac{m_{\tilde{\chi}}}{m_{\tilde{t}}}.
\end{equation*} 
\begin{figure}[t]
\captionsetup{singlelinecheck = false, format= hang, justification=raggedright, font=footnotesize, labelsep=space}
\includegraphics[scale=0.38, trim= 0 0 0 0]{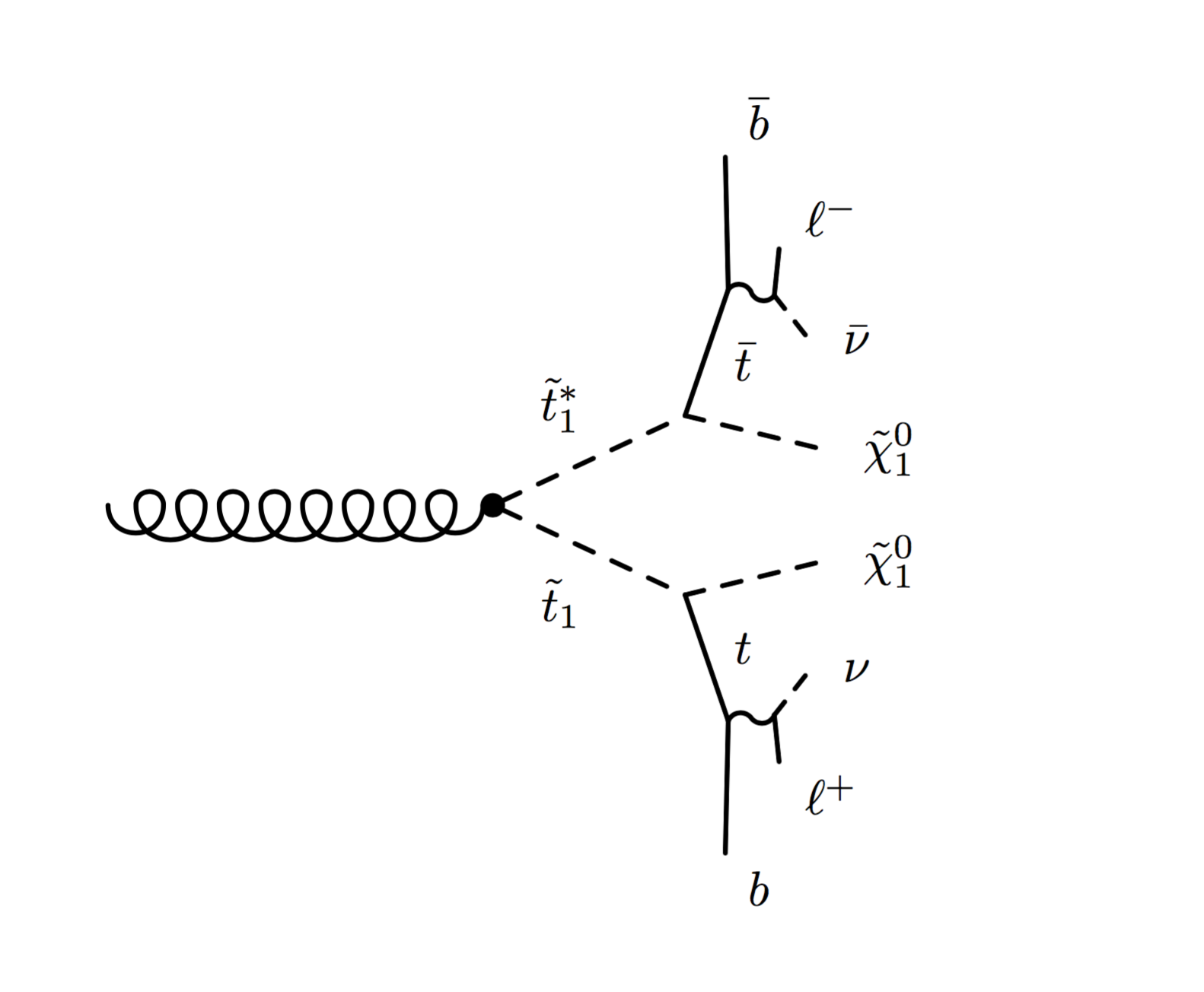}
\caption{The transverse plane topology for the traditional dileptonic stop decay.}
\label{topology}
\end{figure}

For the type of dileptonic stop decays depicted in Fig.~\ref{topology}, which will be referred to as the traditional stop decay from now on, there are 6 mass-shell equations from both top decay chains in an event:
\begin{equation}\label{mass_dilep}
\begin{split}
p_\nu^2=p_{\bar{\nu}}^2 &= 0,\\
(p_{\ell^+}+p_\nu)^2 =(p_{\ell^-}+p_{\bar{\nu}})^2 &= m_{{W}}^2, \\
(p_{\ell^+}+p_\nu+p_{\bar{b}})^2 =(p_{\ell^-}+p_{\bar{\nu}}+p_b)^2&= m_t^2.
\end{split}
\end{equation}
There are two more equations from the transverse momentum conservation, from the components perpendicular and parallel to the $p_T$ of the ISR. 
\begin{equation}\label{met_perp}
\slashed{p}_T^{\perp}=\sum^2_{i=1} {p_{T\nu}^{\perp}}_i, \qquad
\slashed{p}_T^{\parallel}=\sum^2_{i=1} p_{T\nu,i}^{\parallel}+\slashed{p}_{T(\tilde{\chi})},
\end{equation}
where we assumed $\slashed{p}_{T(\tilde{\chi})}^\perp \approx 0$ which is valid in the top corridor.

For such an event we have 8 equations but 9 variables, namely the four-momenta for the two neutrinos and the transverse momentum due to the neutralinos, $\slashed{p}_{T(\tilde{\chi})}$.
Note that only the total transverse momentum of the two neutralinos appears in the kinematic equations but not the individual transverse momentum of each neutralino.
Although we cannot reconstruct all particle momenta for this signal topology, we may ask the question: what is the range of $\slashed{p}_{T(\tilde{\chi})}$ that is compatible with all the kinematic constraints from the 8 equations? This allows us to obtain an upper and a lower bounds for the value of $\bar{R}_M$, which is now adjusted to take into account of the dileptonic nature of the signal:
\begin{equation}
\label{eqn:RMbardefinition}
\bar{R}_M\equiv\frac{\slashed{p}_{T(\tilde{\chi})}}{p_{T(J_{\text{ISR}})}}
\approx\frac{\slashed{p}^{\parallel}_{T}-\sum_i p^{\parallel}_{T\nu,i}}{p_{T(J_{\text{ISR}})}},\quad i=1,2 .
\end{equation} 
Because we do not obtain a single value for $\bar{R}_M$ but just an allowed range, we include a sign in the definition of $\bar{R}_M$ such that it is positive if the (solved) $\slashed{p}_{T(\tilde{\chi})}$ is antiparallel to $p_{T(J_{\text{ISR}})}$ and negative if they are parallel. Effectively, one can include a factor 
\begin{equation}
- \text{sgn} \left(\vec{\slashed{p}}_{T(\tilde{\chi})} \cdot \vec{p}_{T(J_{\text{ISR}})}\right)
\end{equation}
in the $\bar{R}_M$ definition. This is in accordance with the convention that the true $\bar{R}_M$ value is positive for the stop events.

\begin{figure}[th]
\captionsetup{singlelinecheck = false, format= hang, justification=raggedright, font=footnotesize, labelsep=space}
\includegraphics[scale=0.45]{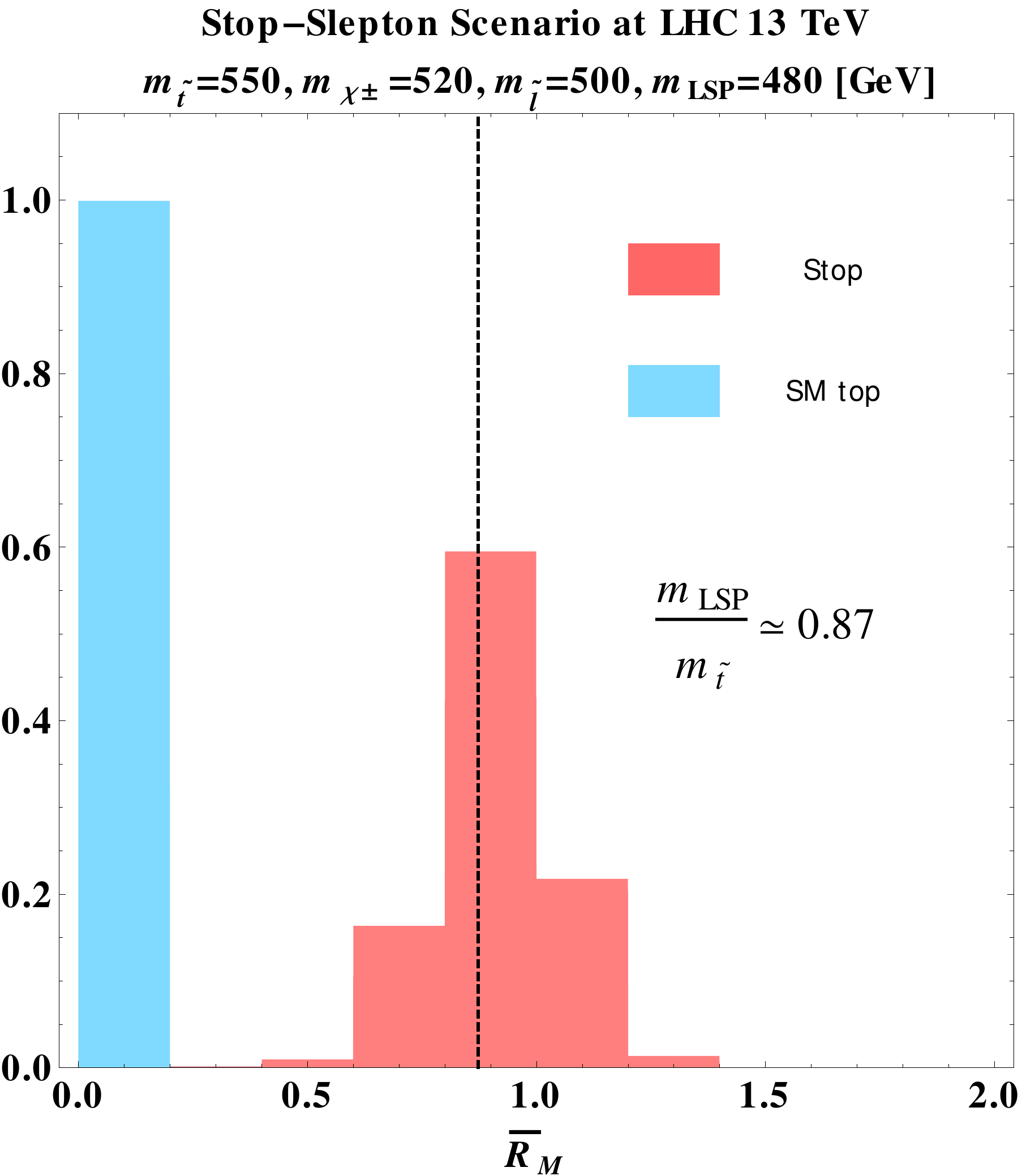}
\caption{The normalized distribution of $\bar{R}_M = \slashed{p}^{\parallel}_{T(\tilde{\chi})}/{p_{T(J_{\text{ISR}})}}$ for the stop-slepton signal with a hard ISR (red) at the parton level for a given spectrum. It is narrowly peaked near the value ${m_{\tilde{\chi}}}/{m_{\tilde{t}}}$. On the other hand, for SM $t\bar{t}$ background, there is no $\slashed{p}_{T(\tilde{\chi})}$ so it is located at 0 (cyan).
}
\label{fig:RMbartrue}
\end{figure}

For the $t\bar{t}$ background where there are no neutralinos, the events should solvable by setting $\slashed{p}_{T(\tilde{\chi})}=0$ in the above equations, assuming that there is no experimental smearing effect. Therefore, the allowed range for $\bar{R}_M$ should contain the point zero, and the upper and lower bounds of $\bar{R}_M$ are expected to converge to zero in the highly boosted (large ISR) regime. Similarly, for the compressed stop events which follow the traditional $t \tilde{\chi}_1^0$ decays, the allowed range is expected to be around its true $\bar{R}_M=\frac{m_{\tilde{\chi}}}{m_{\tilde{t}}}$ value and converges to it in the highly boosted case. Of course, in reality, experimental smearing effects always cause some uncertainties or errors in determining the allowed $\bar{R}_M$ values. As long as the errors introduced by the experimental measurements are small compared to the difference between the theoretical $\bar{R}_M$ value and zero, we expect that the experimentally determined $\bar{R}_M$ variables are useful in distinguishing the traditional stop decay signals from the main background.
On the other hand, for the dileptonic stop decays through sleptons as depicted in Fig.~\ref{fig:feyndiag}, which we will refer to as the stop-slepton decay, there is no unique true value for $\bar{R}_M$ because the kinematic equations do not apply to this topology.
However, If we still define $\bar{R}_M$ for each event to be the ratio between the antiparallel component of $p_T$ of the two neutralinos and the $p_T$ of the ISR, we should get a distribution away from 0 for the stop events. In Fig.~\ref{fig:RMbartrue}, we show such a distribution from an example spectrum at the parton level. One can see that indeed it has a relatively narrow distribution around $m_{\tilde{\chi}}/ m_{\tilde{t}}$. Of course, this $\bar{R}_M$ value is not directly measurable experimentally and only $\bar{R}_{\text{min}}$, $\bar{R}_{\text{max}}$ can be calculated from the experimental observables as its approximations or possible range. Nevertheless, we expect $\bar{R}_{\text{min}}$, $\bar{R}_{\text{max}}$ to have different distributions for the signals and backgrounds, so they may also be used to suppress the background in this case. 
%

In calculating $\bar{R}_{\text{min}}$ and $\bar{R}_{\text{max}}$ for an event with the $2\ell$2$b+\slashed{p}_T+$jet(s) final state, there are some practical issues. First, there are two possible combinations of pairing the $b$-jet and the lepton on the same decay chain. In addition, because the kinematic equations combine into a quartic equation, there could be two disjoint allowed regions for $\bar{R}_M$ in the solution, although it does not happen very often. There is no canonical way to deal with the multiple solutions. In Appendix~\ref{app:technical}, we describe the procedure of solving the kinematic equations and how we define $\bar{R}_{\text{min}}$ and $\bar{R}_{\text{max}}$ for our benchmark studies in more details. Different strategies of choosing $\bar{R}_{\text{min}}$ and $\bar{R}_{\text{max}}$ due to these ambiguities do not have significant effects on the final results.

The SM background events where leptons arise from $W$ decays, such as $t\bar{t}$ or $tW$, tend to have small azimuthal angle separation between the $\slashed{p}_T$ and the light leptons. This can be explored to further suppress the SM dileptonic backgrounds. For single leptonic final states, removing such backgrounds can be done with the transverse mass $M_T$, which has a sharp drop-off around $m_W$ for the SM backgrounds.
For the dileptonic events, there are two leptons in the final states.  A variable called leverage inspired by the transverse mass was defined in Ref.~\cite{Cheng:2016npb} to further reduce the SM backgrounds:
\begin{equation}
L_\ell=\left[\slashed{p}_T
{\sum_i} (1-\cos \Delta \phi_{\ell_i,\slashed{p}_T})\right]/N_\ell,
\end{equation} 
where $N_{\ell}$ is the number of isolated leptons. By definition, if a lepton has a low azimuthal angle separation $\Delta\phi_{\ell_i,\slashed{p}_T} \sim 0$ from the $\slashed{p}_T$ direction, its contribution to $L_\ell$ is small. As the stop signal events tend to have larger MET and larger lepton-$\slashed{p}_T$ azimuthal separations due to the $\tilde{\chi}_1^0$'s in the final states, a minimum cut on $L_\ell$ can effectively suppress the SM backgrounds and enhance the signal significance.

\section{Compressed Stop in Dileptonic Searches}
\label{Section:Bench}
In this section, we perform some more detailed collider studies for both the stop-slepton decay case and the traditional stop decay case with compressed spectra. 
We use MadGraph 5~\cite{Alwall:2011uj} and Pythia 6~\cite{Sjostrand:2006za} to generate both the background and the signal events. MLM matching scheme~\cite{Mangano:2002ea} is applied for both the $t\bar{t}$ background and the SUSY signal production in order to prevent double-counting between the matrix elements and the parton shower. The detector simulation is performed by Delphes 3~\cite{deFavereau:2013fsa}. For the signals, the production cross section is normalized to 13 TeV NLO+NLL results~\cite{Borschensky:2014cia}. The $b$-tagging efficiency is taken to be a universal $70\%$ with an overall light-flavor mis-tag rate of 1.5$\%$.

We expect that the SM $t\bar{t}$(+ jets) production to be the dominant $b\bar{b}\ell^+\ell^- +\slashed{p}_T$ background. The NLO cross section given by MadGraph5 aMC@NLO~\cite{Alwall:2014hca} is 70.9~pb. Compared to the LO result this corresponds to a $K$-factor about 1.5. This is also consistent with the dedicated NLO calculations of the $t\bar{t}+$ jets cross sections~\cite{Hoeche:2014qda,Hoche:2016elu}. Therefore, we multiply the number of events generated at LO by a $K$-factor of 1.5 to match to the NLO calculation after the preliminary selections described below.
Besides $t\bar{t}$, the backgrounds from $tW$ and $t\bar{t}+W/Z$ production were simulated with LO cross section 2.68~pb and 0.94~pb, since their impacts on the total background are small. All backgrounds were generated with the corresponding integrated luminosities greater than 300 $\ifb$. Other SM backgrounds, such as diboson production or $Z+$jets, have small cross sections or low signal efficiencies. Consequently we ignore these background processes for the rest of our discussion.

For our benchmark studies, all the events must satisfy the preliminary selections as described below. All the events are required to have two $b$-tagged jets with $p_T > 25$~GeV, two light leptons with $p_T > 20$~GeV.\footnote{For smaller mass splittings, one may lower the $p_T$ cut to increase the signal efficiency. However, there will also be more backgrounds and eventually one runs out of the sensitivity for very compressed spectra.}
Since our analysis relies on a hard ISR jet, we require that the hardest non-$b$-tagged jet has $p_T>150$~GeV.  To take into account the cases where there are more than one ISR jets, we define ${p_T}_{\text{ISR}}$ to be the vector sum of the three leading non-$b$-tagged jets. We also require that MET$>150$~GeV, since the signal is expected to have a substantial amount of missing transverse momentum.
Furthermore, to reduce $Z\to \ell\ell$ and $\tau$ related backgrounds, we veto all events with a $\tau$-tagged jet or opposite-sign same-flavor lepton pairs with $M_{\ell\ell}\in[m_Z-10,m_Z+10]$.

\subsection{Stop-Chargino-Slepton Scenario}

For the stop $\to$ chargino $\to$ slepton type of decay, we take a simplified model approach. Assuming that $\mchar$ and $\mneutwo$ are degenerate, so are $\mslep$ and $\msneu$. The simplified model is therefore characterized by four masses: $\mtt$, $\mchar$, $\mslep$ and $\mlsp$. To suppress the direct $\tt \to t\lsp$ decay, the $\mtt-\mlsp$ gap is chosen to be smaller than $m_t$. We assume that all stops decay through the chargino and the (first two generation) sleptons with a 50\% branching ratio through either the charged sleptons or the sneutrinos. Since we are interested in the ``compressed'' spectrum, $\mtt-\mchar$ and $\mchar-\mlsp$ are chosen to be small, so that the searches based on the $\mathrm{M_{T2}}$ type of variables~\cite{CMS:2017qjo} are ineffective. This allows us to explore the usefulness of the $\bar{R}_M$ type of variables. 
In this subsection we show detailed analyses on two signal benchmarks: BMP1$=(\mtt=550$~GeV, $\mchar=520$~GeV, $\mslep=500$~GeV, $\mlsp=480$~GeV) and BMP2$=(\mtt=670$~GeV, $\mchar=640$~GeV, $\mslep=600$~GeV, $\mlsp=560$~GeV). 
An expected search reach for $\mtt-\mchar=30$~GeV in the $\mchar$--$\mlsp$ plane is presented later.

\begin{figure}[th]
\captionsetup{singlelinecheck = false, format= hang, justification=raggedright, font=footnotesize, labelsep=space}
\includegraphics[scale=0.45]{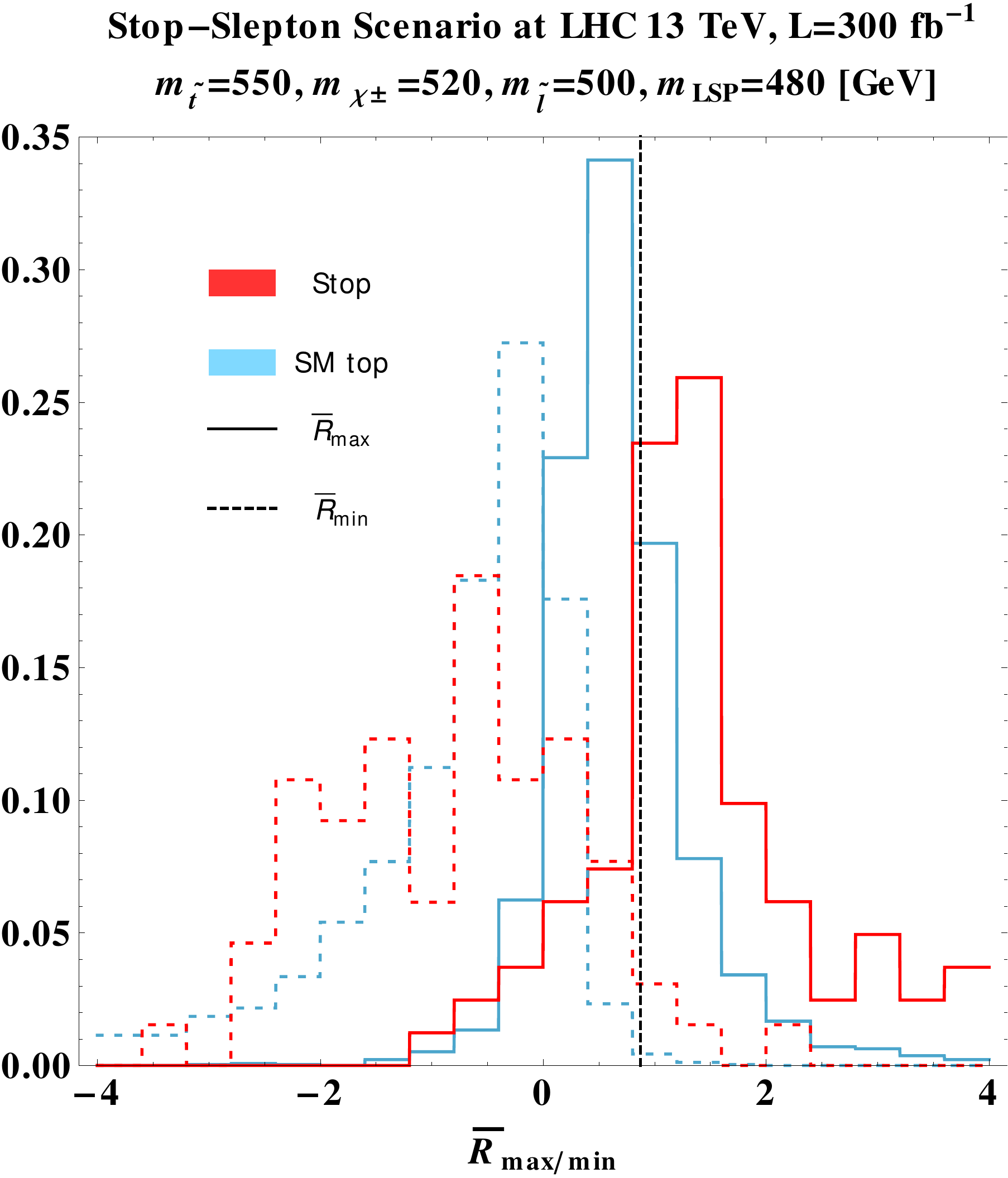}
\caption{The normalized one-dimensional distributions of $\bar{R}_{\text{max}}$ and $\bar{R}_{\text{min}}$ for the stop signal events and SM $t\bar{t}$ background events for the BMP1 spectrum.}
\label{fig:1D}
\end{figure}
The (normalized) distributions of $\bar{R}_{\text{max}}$ and $\bar{R}_{\text{min}}$ for the stop signal events and SM $t\bar{t}$ background events for BMP1 are shown in Fig.~\ref{fig:1D}. For this benchmark. $\bar{R}_{\text{max}}$ tends to be larger for the signal than the background, while it is less clear for  $\bar{R}_{\text{min}}$. However, these variables can be correlated so it is useful to examine the two-dimensional distributions to further enhance the discriminating power.
\begin{figure}[t]
\captionsetup{singlelinecheck = false, format= hang, justification=raggedright, font=footnotesize, labelsep=space}
\begin{subfigure}[b]{0.45\textwidth}
\includegraphics[width=7cm]{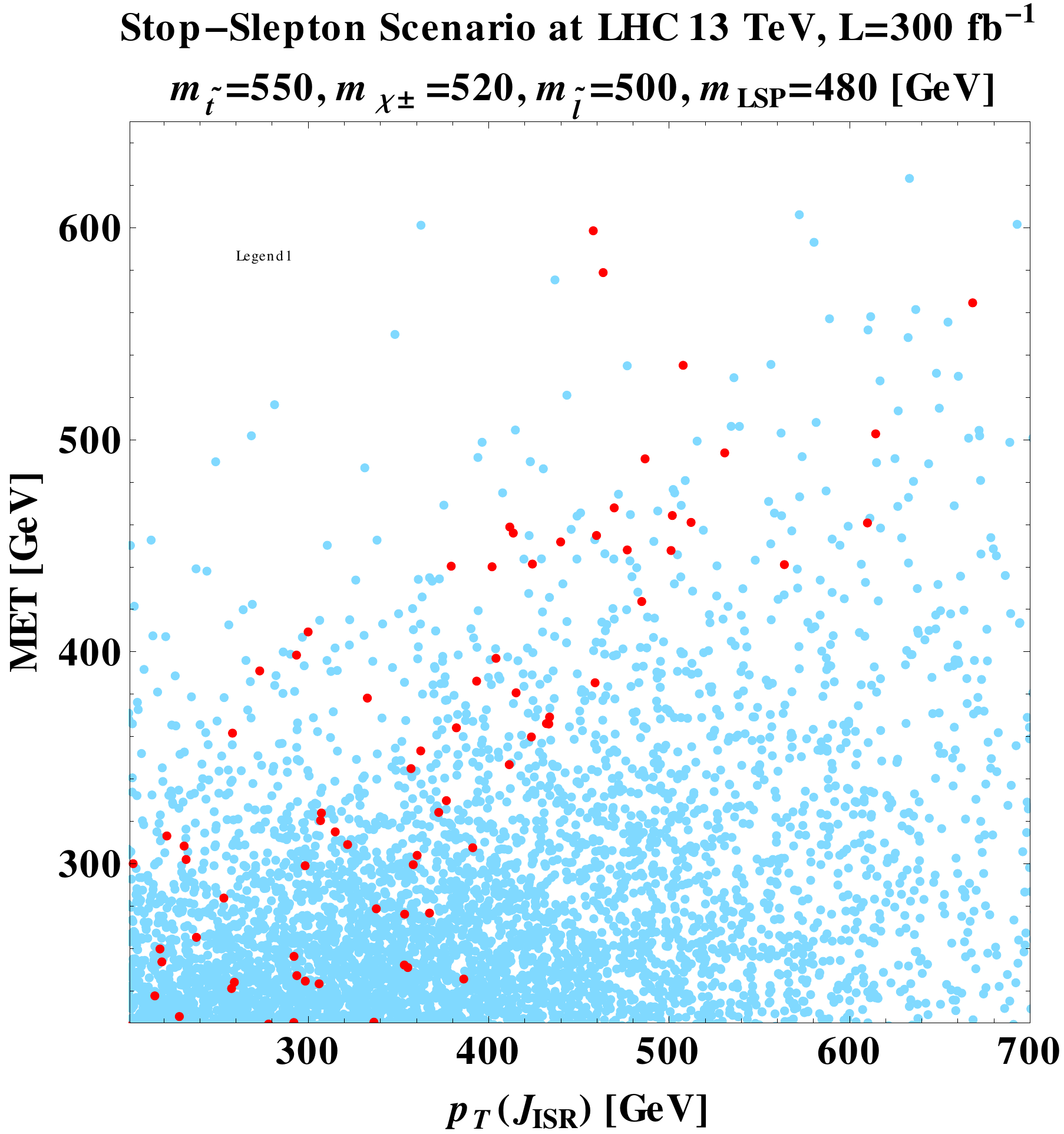}
\end{subfigure}
\begin{subfigure}[b]{0.45\textwidth}
\includegraphics[width=7cm]{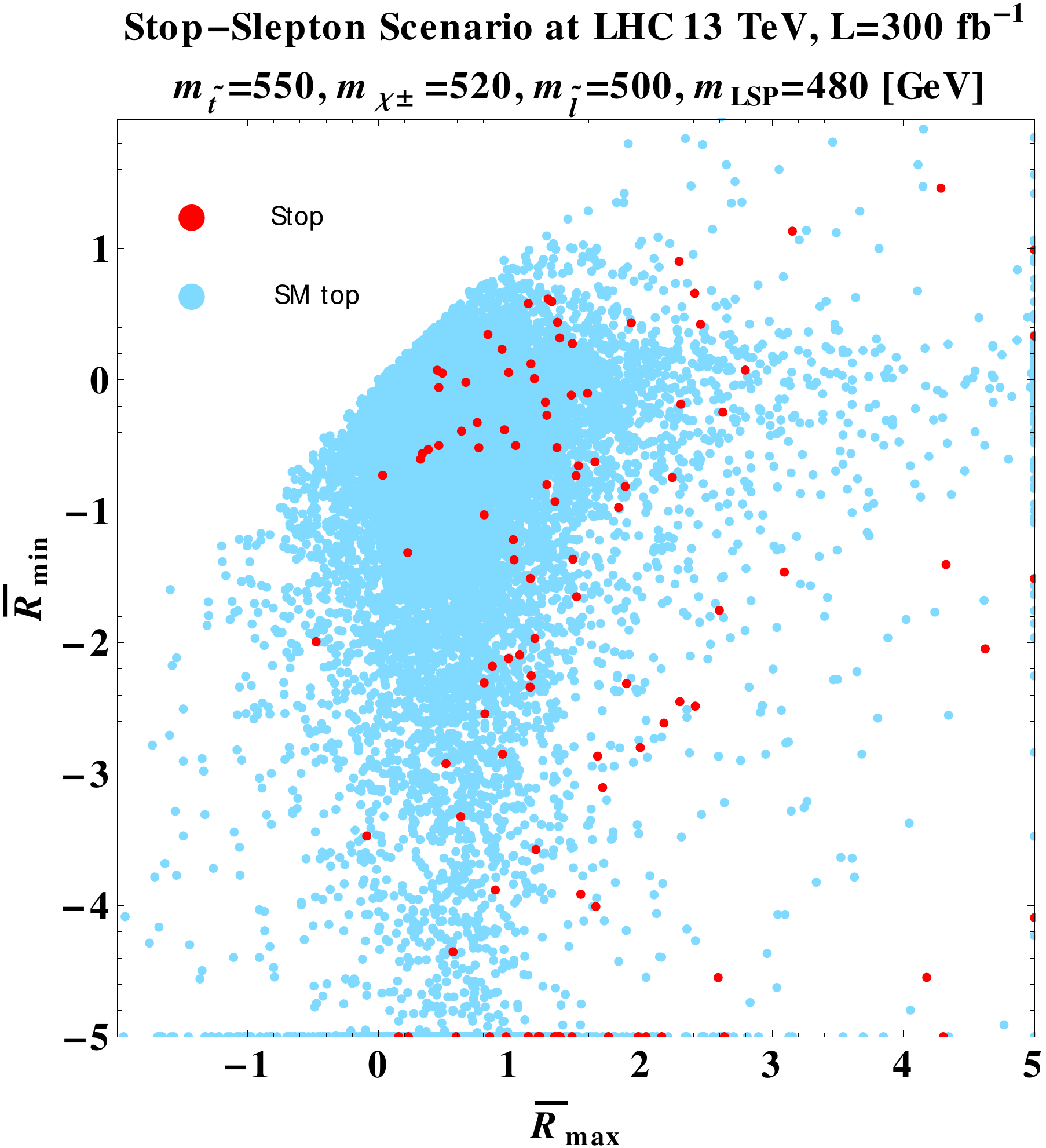}
\end{subfigure}
\begin{subfigure}[b]{0.45\textwidth}
\includegraphics[width=8cm]{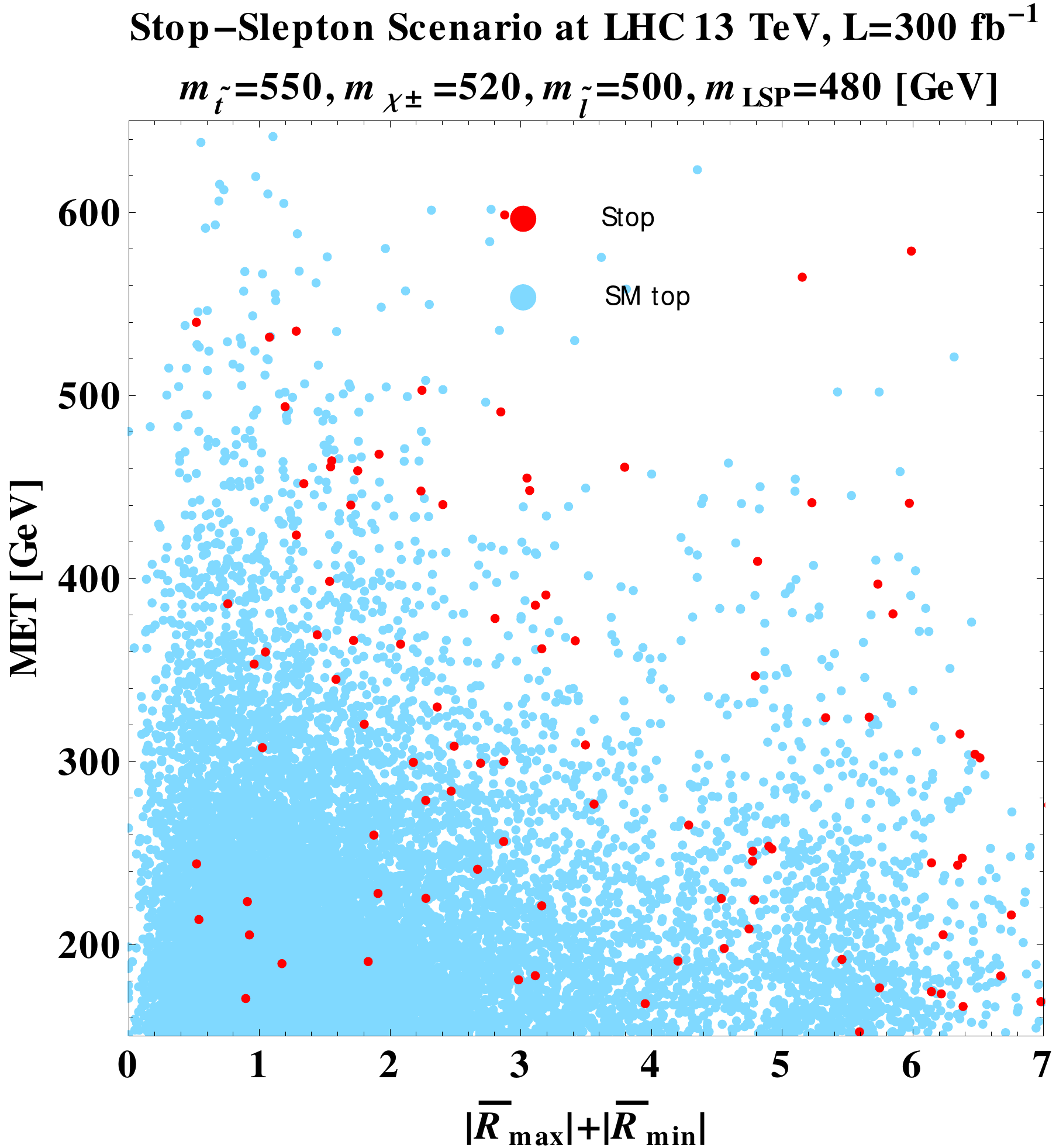}
\end{subfigure}
\caption{Two-dimensional distributions of the signal (red points) vs. $t\bar{t}$ background (blue points) for the benchmark point BMP1 ($\mtt / \mchar / \mslep / \mlsp = 550/520/500/480$~GeV). Both the signal and the background are weighted to 300 fb$^{-1}$.}
\label{fig:first}
\end{figure}
In Fig.~\ref{fig:first}, we plot a series of two-dimensional distributions for the benchmark BMP1 signals and the $t\bar{t}$ background after the preliminary selections. It can be clearly seen that the signal events tend to have a larger MET, especially for a larger ${p_T}_{\text{ISR}}$. We also notice that in the $\Rmbmax$ vs. $\Rmbmin$ plane, signals have a different feature of the distribution compared to the backgrounds.\footnote{One may notice that a small fraction of events (especially for backgrounds) has both $\Rmbmax$ and $\Rmbmin$ negative. This could due to a number of effects, like smearing, wrong combinations, and inaccurate identification of the ISR system. If the measured missing $p_T$ antiparallel to the ISR is smaller than the true value (or the reconstructed ISR momentum is bigger than the true value), one may require the ``neutralino momentum'' to be in the same direction as the ISR to solve the constraint equations, resulting in negative values of $\Rmb$.} 
As discussed in Sec.~\ref{sec:kinematics}, the SM $t\bar{t}$ events are expected to have $\bar{R}_M$ close to 0. This can be recognized in the $|\Rmbmax|+|\Rmbmin|$ distribution, as shown in the bottom panel of Fig.~\ref{fig:first}.
The $|\Rmbmax|+|\Rmbmin|$ distribution for the $t\bar{t}$ background is smaller than that of the signal, especially when MET increases.

To obtain a better search reach, one can divide the events into many signal regions based on these kinematic variables and perform a multi-variate analysis. However, to get a good intuition on how the kinematic variables discriminate the signal and the backgrounds, we perform a simpler cut-and-count analysis to select a few signal regions in this theoretical study.
First, we require ${p_{T}}_{\text{ISR}}\geqslant 200$~ GeV and MET$> 225$~GeV. Also, since the benchmark spectra focused in this study are compressed, we impose a cut on the leading lepton $p_T < 100$~GeV for all events.
Then we apply cuts on the pre-selected events and divide them into the following three exclusive signal regions (SR) based on their MET: 
\begin{itemize}

\item for events with MET$\geqslant$ 550~GeV,  $|\Rmbmax| +|\Rmbmin|>1$ is required (SRH);

\item for events with MET $\in(325,500)$~GeV, $|\Rmbmax| +|\Rmbmin|  > 4.6 -\frac{\text{MET}}{150~\text{GeV}}$ and Leverage $L_\ell>$ 80~GeV are required (SRM);

\item for events with MET $\in(225,325)$~GeV,$|\Rmbmax| +|\Rmbmin|>2.5$, $ \Rmbmax >1 $ and $ \Rmbmin<-1 $, $\frac{\text{MET}}{{p_T}_{\text{ISR}}} > 0.75$, Leverage $L_\ell>$ 80~GeV are required (SRL).

\end{itemize}
The selection criteria for the signal regions are motivated by that event distributions shown in Fig.~\ref{fig:first}. For large MET, the $|\Rmbmax| +|\Rmbmin|$ values of the background events stay small and hence we impose a looser cut on $|\Rmbmax| +|\Rmbmin|$. On the other hand, for small MET, more stringent cuts on $|\Rmbmax| +|\Rmbmin|$ and/or other variables are needed to effectively reduce the background. Consequently, SRM and SRL  are supplemented with a cut $L_\ell>$ 80~GeV besides the $|\Rmbmax| +|\Rmbmin|$ requirements. The optimal selection criteria in principle can depend on the benchmark point. However, to have a more general analysis without relying on small details, we use the same 3 signal regions for all points that we studied and we found that they can improve the signal significances effectively.

The cut flow and the number of events passing the cuts for two benchmark points and the SM backgrounds, normalized to an integrated luminosity of 300~fb$^{-1}$ are shown in Table~\ref{tab:1}.
As expected, the SM $t\bar{t}$ is the dominant background for all signal regions. 
\begin{center}
\begin{table}[ht]
\captionsetup{singlelinecheck = false, format= hang, justification=raggedright, font=footnotesize, labelsep=space}
\begin{small}
\begin{tabular}{l|ccccc|ccc}
         & Initial & Pre-selection & $\,\,\,{p_{T}}_{\text{ISR}}\geqslant 200$ & $\,\,\,p_{T\ell_1} \leqslant 100$ & MET$> 225$ & SRL &SRM & SRH \\ 
\hline\hline
BMP1 & 8.9$\times 10^4$ & 167 & 144 & 144 &  115  & 24.5 & 27.5 & 10.7 \\
\hline
BMP2 & 2.6$\times 10^4$ & 145 & 121 & 116 &  96.2  & 4.8 & 19.6 & 11.5 \\
\hline
\hline
SM $t\bar{t}j$ & $ 2.1\times 10^7$ &  2.12 $\times 10^5$ &  1.68 $\times 10^5$ & 1.18 $\times 10^5$  & 3.77 $\times 10^3$ & 126 & 129 & 20.5 \\
SM $tWj$ & 8.1 $\times 10^5$ & 627 & 459 & 293 & 124 & 1.7 & 2.6 & 0.9 \\
SM $t\bar{t}V$ & 2.8 $\times 10^5$ & 166 & 121 & 78 & 32 &0.7 & 0.7 & 1.4 \\
\hline\hline
SM total & $2.2\times 10^7$ & 1.47$\times 10^5$&  1.17$\times 10^5$& 8.21$\times 10^4$ &2.76$\times 10^3$ & 129  & 132  & 22.8 \\
\end{tabular} 
\end{small}
\caption{The cut flow for the stop-slepton scenario, assuming an integrated luminosity of  300 fb$^{-1}$. The preliminary cuts are described in the beginning of Sec.~\ref{Section:Bench}. The benchmark BMP1 has the spectrum $\mtt/ \mchar/\mslep/ \mlsp= 550/520/500/480$~GeV, while the corresponding spectrum for BMP2 is 670/640/600/560~GeV. The last three columns are the low, medium, high MET signal regions as defined in the text.}
\label{tab:1}
\end{table}
\end{center}

To calculate the signal significances for the benchmark models, we use the likelihood method with the assumption
that the overall number of background events in each signal region respects the normal distribution with a fractional uncertainty $\sigma_{B}\propto B$. The likelihood is defined to be 
\begin{equation}
Q=\frac{\int \mathcal{L}(S+B,S+B')P(B')dB'}{\int \mathcal{L}(S+B,B')P(B')dB'},
\end{equation}
where $S$ and $B$ are corresponding numbers of signal and background events, $\mathcal{L}(x,\mu)=\frac{\mu^{x}e^{-\mu}}{x!}$, and $P(B)$ is the normalized normal distribution with the mean $B$ and a standard deviation $\sigma_B$. The final significance from this method is simply given by $\sqrt{2\log(Q)}$. 
For the case with no systematic error, $\sigma_B=0$, this equation simply reduces to the standard formula~\cite{Cowan:2010js}:
\begin{equation}
\sigma = \sqrt{2\left[(S+B)\log\left( \frac{S+B}{B} \right)-S\right]}.
\label{eq:significance}
\end{equation}
Assuming the statistical fluctuations are independent in different SR's, the overall significance is obtained by combining those of the 3 SR's in quadrature. For BMP1, we get a significance of $2.8\sigma\,(3.8\sigma)$  for 300~$\ifb$ with (without) a 10$\%$ background uncertainty. For BMP2, we get a $2.3\sigma$ ($2.8\sigma$) significance. As shown in the cut flows for BMP2, the contribution from the higher MET bins (SRH, SRM) become more important for larger signal masses. Since the number of background events significantly decreases in SRH, the background uncertainty affects the significance far less for BMP2 compared to BMP1.

One could ask how much the new $\Rmb$ variables really help the stop search in this case, given that the signal and background distributions already look different in the standard variables such as $\slashed{p}_T$ and ${p_T}_{\text{ISR}}$. In Appendix~\ref{app:validation} we compare the analyses with and without the $\Rmb$ variables by dividing the variable space into the same number of signal regions and show that the inclusion of $\Rmb$ variables does substantially improve the signal significance.

To explore the search reach of the stop-slepton decay case, we repeat the analysis for a range of different spectra. 
To simplify the study, we follow the assumptions as adopted in the benchmark study. We take $\mchar$ and $\mlsp$ as the free parameters and fix $\mtt-\mchar=$30~GeV, and $\mslep=(\mchar+\mlsp)/2$. The $\mtt-\mchar=$30~GeV is a moderate choice where we still have a good $b$-jet tagging efficiency. For smaller mass differences the $b$-jets become too soft and signal efficiency deteriorates significantly. A comparison of overall signal efficiencies for several mass differences is listed in Table~\ref{tab:massgap}.
\begin{table}[ht]
\captionsetup{singlelinecheck = false, format= hang, justification=raggedright, font=footnotesize, labelsep=space}
\begin{tabular}{c|c|c|c}
$\mtt-\mchar$/GeV & 15 & 30 & 60 \\ 
\hline\hline 
Overall efficiency & $5.0\times 10^{-5}$ & $7.1\times 10^{-4}$ & $1.3\times 10^{-3}$ 
\end{tabular} 
\caption{{The overall signal efficiencies for $\mtt=550$~GeV with varied $\mtt-\mchar$. We also fix $\mchar-\mslep$ and $\mslep-\mlsp$ to be 20~GeV. Here the overall signal efficiency includes all 3 SRs. }}
\label{tab:massgap}
\end{table}

We then calculate the significances for points in the plane of $\mchar$ vs.~$\mlsp$ and obtain the 2$\sigma$ exclusion reach for an integrated luminosity of 300~fb$^{-1}$. The result is shown in Fig.~\ref{fig:slepton_limit}.
The most stringent constraint for this kind of spectrum comes from the $\tilde{\chi}_1^{\pm}-\tilde{\chi}_2^{0}$ production, which gives 3 lepton or same-sign dilepton final states. The current limit on the chargino mass can be up to 1150~GeV for a light LSP mass~\cite{CMS:2017fdz,ATLAS:2017uun}. However, the mass reach for a compressed spectrum is limited. The current CMS exclusion limit for $\mslep=(\mchar+\mlsp)/2$ is also plotted in Fig.~\ref{fig:slepton_limit}. 
For fixed $\mchar$ and $\mlsp$ masses, the bound has mild dependence on $\mslep$, since the typical lepton momentum depends on the mass splittings $\mchar-\mslep$ and $\mslep-\mlsp$. For instance, if $\mchar-\mslep \ll \mslep-\mlsp$, the lepton from the $\tilde{\chi}^{\pm}_1 \to \tilde{\nu}^{(*)} + \ell^{\pm}$ decay chain would become soft, which leads to lower signal efficiency. On the other hand, the lepton from the other decay chain, namely $\tilde{\ell}^{\pm} + \tilde{\chi}_1^0 + \ell^{\pm}$, would have a larger momentum and hence a higher efficiency. Such a compensation effect between the two decay chains results in only mild changes in the exclusion limit when varying $\mslep$.  For instance, in the current LHC chargino/slepton searches~\cite{CMS:2017fdz}, the exclusion limit for $\mlsp$ with $\mchar=500$~GeV and $\mslep=(\mchar+\mlsp)/2$ is $\sim 370$~GeV. The limit changes to $\sim 350$~GeV for both  $\mslep = 0.95 \mchar + 0.05 \mlsp$ and $\mslep = 0.05 \mchar + 0.95 \mlsp$ cases. For our benchmark parameters, $\mtt=550$~GeV with $\mchar=\mneutwo=$520~GeV and $\mlsp=480$~GeV, our analysis shows that the signal significance for $\mchar-\mslep=5$~GeV is $\sim 60\%$ of that of $\mchar-\mslep=20$~GeV.

\begin{figure}[t]
\captionsetup{singlelinecheck = false, format= hang, justification=raggedright, font=footnotesize, labelsep=space}
\includegraphics[width=12cm]{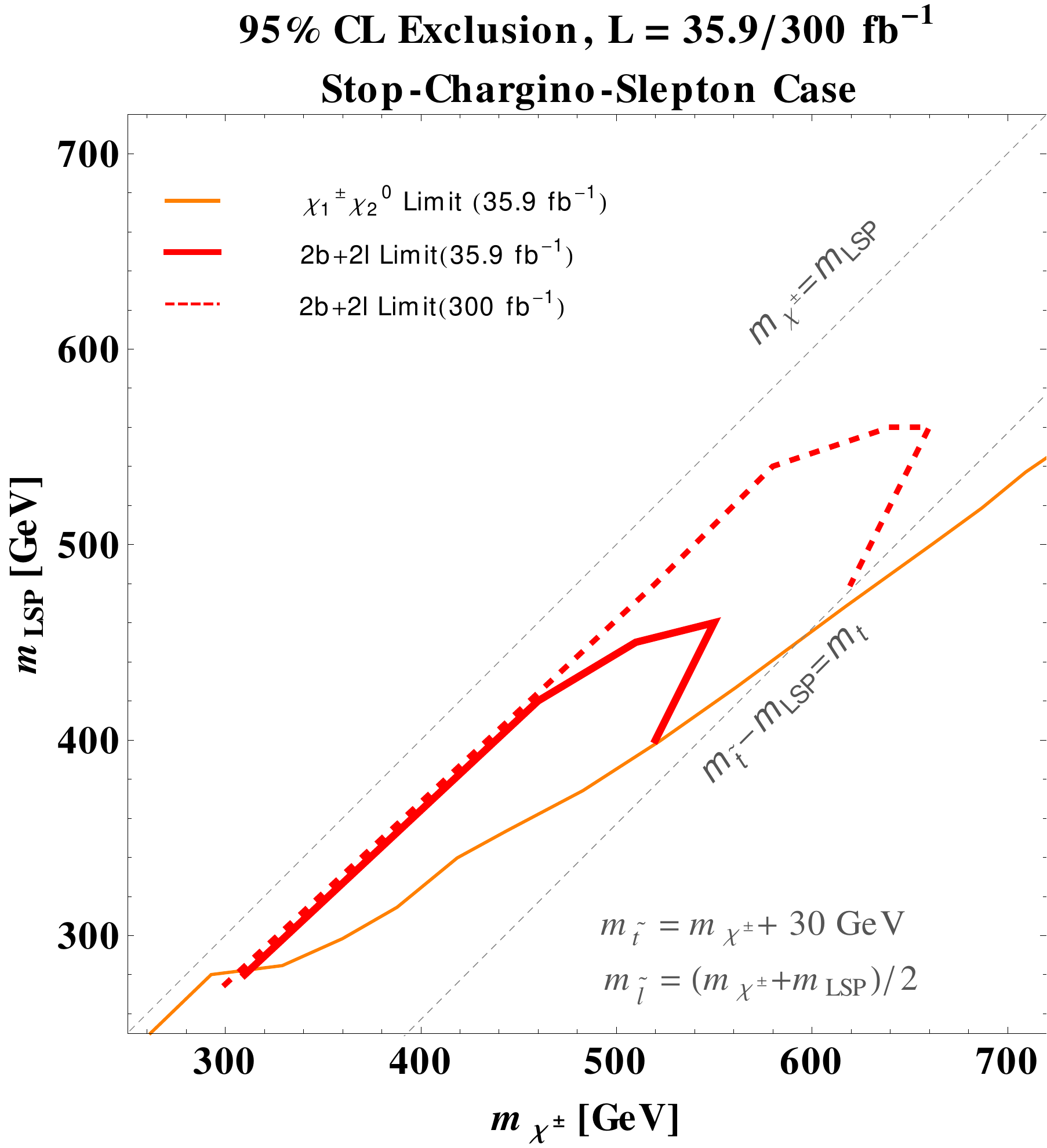}
\caption{The (2$\sigma$) exclusion search limits from the dileptonic stop-slepton analyses with 36 fb$^{-1}$ (solid red) and 300 fb$^{-1}$ (dashed red) integrated luminosities.  The stop masses are chosen to be 30~GeV heavier than the chargino, and the slepton masses are chosen to be the average of the chargino and the LSP mass. The orange curve is the CMS 35.9 $\ifb$ exclusion limit, coming from the same-sign dilepton or trilepton $\chi^0\chi^{\pm}$ search results~\cite{CMS:2017fdz}.
}
\label{fig:slepton_limit}
\end{figure}

One can see that the dileptonic stop search in the stop-slepton decay case can probe the parameter space which is not excluded by the current experimental limits. One should note that our search is based on the stop pair production while the exclusion limit on Fig.~\ref{fig:slepton_limit} comes from $\tilde{\chi}_1^{\pm}-\tilde{\chi}_2^{0}$ production which is independent of the stop mass. 
Fig.~\ref{fig:slepton_limit} should not be viewed as a comparison of the powers of the two different searches, but it demonstrates that the dilepton stop search discussed in this paper can probe parameter regions that can not be reached by just the chargino-neutralino searches.
With a relatively light stop, the stop pair production has the advantage of larger production cross section. Although introducing $b$ jets in our final state would increase background events, the kinematic variables introduced in this work allow us to better handle the backgrounds. We should also point out that our projected reach is based on the simple cut-and-count analysis. It might be further improved with a more sophisticated multivariate analysis.

\subsection{Traditional Stop Decays with a Compressed Spectrum}

In this subsection we turn to the compressed stop searches with the traditional $t\lsp$ decay into the $b\bar{b}\ell^+\ell^- +\slashed{p}_T$ final states. The all hadronic and semileptonic channels have obtained quite strong limits in the compressed region with the help of the hard ISR and the $R_M$ variable~\cite{ATLAS:2017kyf,CMS:2017vbf,CMS:2017arv,ATLAS:2017msx}. Stop mass up to 590 GeV has been excluded along the top corridor~\cite{ATLAS:2017kyf}. The dileptonic channel suffers from the small branching ratio. In order not to have too few signal events, we choose the point $\mtt$=600~GeV and $\mlsp$=427~GeV just beyond the current limit for our analysis.
In contrast to the stop-chargino-slepton scenario, $\bar{R}_M$ also has a physical interpretation for the signals in this case, and $\bar{R}_M$ would be close to $\frac{\mlsp}{\mtt}$ if the signals are very boosted. Therefore, in addition to the pre-selection criteria as described in the beginning of this section, we trigger on events with ${p_T}_{\text{ISR}} > 200 $ GeV instead of ${p_T}>150$~GeV for a single jet and $\Delta\phi_{j_{\text{ISR}},\slashed{p}_T}>2$.
\begin{figure}[t]
\captionsetup{singlelinecheck = false, format= hang, justification=raggedright, font=footnotesize, labelsep=space}
\includegraphics[width=9cm]{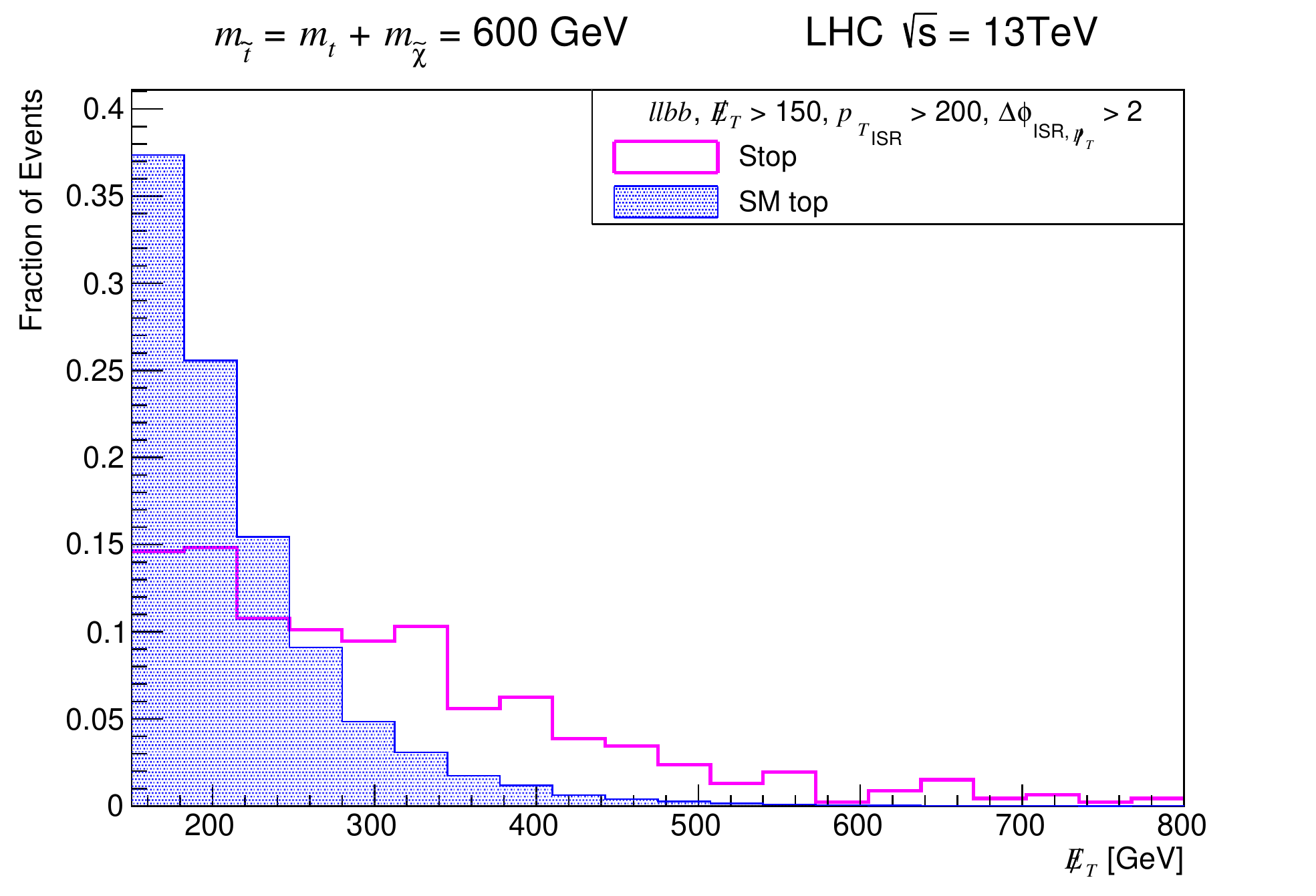}
\caption{Distributions of MET for signal and background 
after the preselections as described in the text. The distributions are normalized to 1.}
\label{kinematics}
\end{figure}
Fig.~\ref{kinematics} shows the MET distribution for signal and backgrounds after the preselections. 
Based on this, we require that MET $>350$ GeV.

In this case, $\Rmbmax$ and $\Rmbmin$ are expected to be around the true $\Rmb$ value which is larger than 0 for the signal, hence we scan the $\Rmbmax$ and $\Rmbmin$ in the range from 0 to 2.  
Fig.~\ref{rm_2d} shows the two dimensional distribution of $\Rmbmax$ and $\Rmbmin$ for the signal and the background after the MET cut. 
Clearly, the background events aggregate near the region where $\Rmbmin=0$. Based on this, we apply the final selection: $\Rmbmin >0.4$ and $\Rmbmax>0.7$. This gives us a significance $\sim1.7\sigma$ at 300 fb$^{-1}$ without taking into account of any background uncertainty.

\begin{figure}[t]
\captionsetup{singlelinecheck = false, format= hang, justification=raggedright, font=footnotesize, labelsep=space}
\includegraphics[width=9cm]{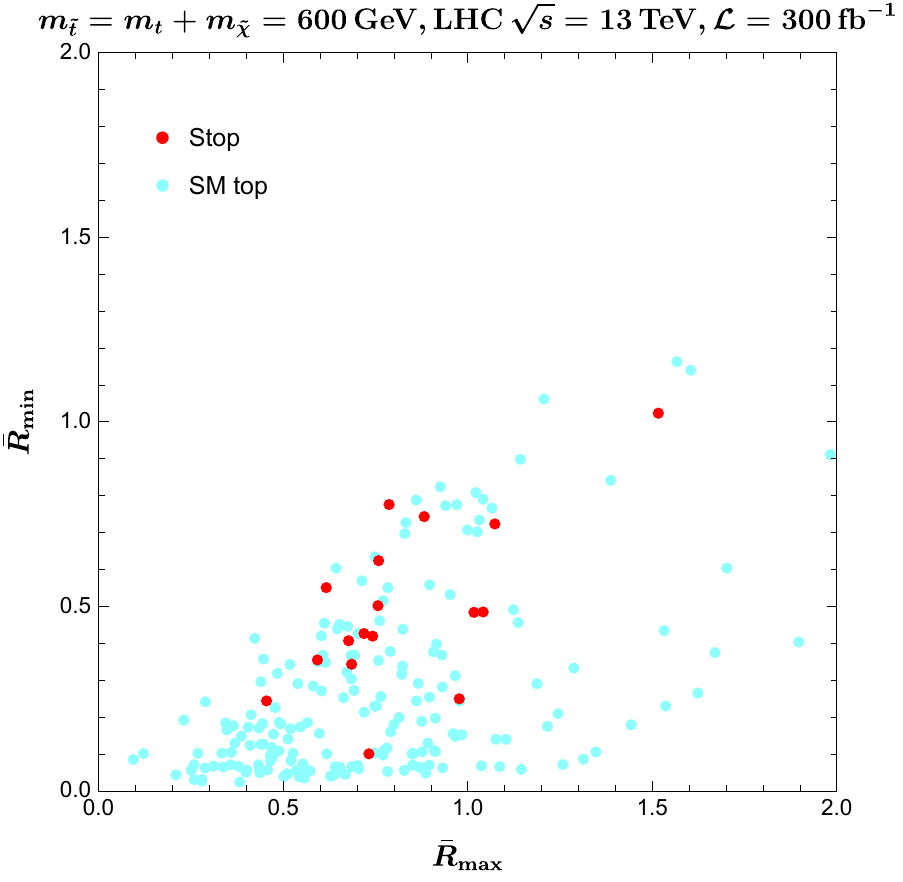}
\caption{The distributions of the signal and the background events after the preselections and the MET cut, projected onto the $\Rmbmax$ vs. $\Rmbmin$ plane.}
\label{rm_2d}
\end{figure}

In Table~\ref{tab:2} we compare the dilepton result (Di) with the semileptonic (Semi) analysis from Ref.~\cite{Cheng:2016mcw} for the chosen benchmark point.  
In the Semi analysis, the preselections require at least 4 jets, 1 or 2 $b$-tagged jets, exactly one light-flavor lepton, MET$>100$~GeV, $\Delta\phi_{j_{\text{ISR}},\slashed{p}_T}>2$ and that the hardest jet (ISR) has $p_T\geqslant 200$~GeV. Afterwards the following cuts were applied: MET$>200$~GeV, $\Delta\phi_{\ell,\slashed{p}_T}>0.9$ and ${p_T}_{\text{ISR}}\geqslant 475$~GeV, which is grouped under ``Other cuts'' in Table~\ref{tab:2}. The  $\bar{R}_M$ cuts select $\bar{R}_M$ within the range ($\frac{\mlsp}{\mtt}-0.15$, 1).

As one can see from Table~\ref{tab:2}, for the chosen benchmark point, the strength of the dileptonic channel based on the simple analysis is slightly weaker than but not far from that of the semileptonic channel. 
The dileptonic channel has fewer signal events but also less background.
We also repeat the dilepton analyses for different mass points along the line $\mtt-\mlsp=m_t$. Due to the limited statistics, the $\Rmbmax$ or $\Rmbmin$ does not show a prominent peak for the signal events. Therefore, they are used more as a tool to identify the $t\bar{t}$ background. We define our signal region to consist of MET$>350$ GeV, $\Rmbmin >0.4$ and $\Rmbmax>0.7$.
The significances of different $\mtt$ are listed in Table~\ref{tab:3}. 
 
Since these numbers are for a future integrated luminosity of 300 fb$^{-1}$, they are not expected to compete with the projected all-hadronic analysis.  It is possible that the significances can be improved with more sophisticated analyses, although the improvement may be limited if the signal is statistically limited as in the case of the dileptonic decay. Nevertheless, analyses based on different final states  provide complementary probes of various stop decay channels. 

\begin{table}[ht]
\captionsetup{singlelinecheck = false, format= hang, justification=raggedright, font=footnotesize, labelsep=space}
\begin{tabular}{l |ccccc}
 & Initial &Pre-selections& \,Other cuts &$\,\,\,\bar{R}_M$cuts\\ 
\hline 
$\tt_1 \tt_1^*j$ (Di) & 5.4$\times 10^3$ & 69 & 27 &12 \\ 
\hline  
SM $t\bar{t}j$(Di) & 2.1$\times 10^7$ & 1.86$\times 10^4$ & 573 & 45\\ 
\hline 
\hline
$\tt_1 \tt_1^*j$ (Semi) & 2.3$\times 10^4$ &597 & 36 & 24 \\ 
\hline 
SM $t\bar{t}j$ (Semi) & 1.2$\times 10^8$  &  5.82$\times 10^5$ & 327& 15\\ 
SM $t\bar{t}j$ (Di) & 2.1$\times 10^7$ &  1.24$\times 10^5$ & 352& 103 \\ 
\end{tabular} 
\caption{Dileptonic (Di) and the semileptonic (Semi)  analysis of the compressed stop pair production, assuming an integrated luminosity of 300 fb$^{-1}$. The main background comes from $t\bar{t}$ pair production. For both cases, the signal benchmark has $\mtt$=600~GeV and $\mlsp$=427~GeV.  
The Pre-selections for Semi and Di analyses are different, since they focus on completely different final states. The details are described in the text above.
}
\label{tab:2}
\end{table}

\begin{table}[ht]
\captionsetup{singlelinecheck = false, format= hang, justification=raggedright, font=footnotesize, labelsep=space}
\begin{tabular}{c |c |c |c |c |c |c |c |c  }
$\mtt$ (GeV) &400 &450& 500 &550 &600&650&700&750\\ 
\hline
\hline
300 fb$^{-1}$ limit  &3.4 (2.9)$\sigma$ &3.2 (2.7)$\sigma$ &2.1 (1.8)$\sigma$ &1.6 (1.4)$\sigma$  &1.7 (1.5)$\sigma$&1.9 (1.6)$\sigma$ &1.3 (1.1)$\sigma$ &1.0 (0.9)$\sigma$\\ 
\end{tabular} 
\caption{Significances achieved by the traditional dileptonic stop decays along the line $\mtt-\mlsp=m_t$, assuming an integrated luminosity of 300 fb$^{-1}$. The signal region is chosen to be MET$>350$ GeV, $\Rmbmin >0.4$ and $\Rmbmax>0.7$. Numbers in the parentheses include an assumed 10\% systematic uncertainty.}
\label{tab:3}
\end{table}

\section{Conclusions}
\label{sec:conclusions}

The stops are the most relevant particles for the naturalness of the SM if SUSY is the solution to the hierarchy problem. The experimental verification of whether they exist is of undisputed importance in testing SUSY as a possible new symmetry principle of the universe and our understanding of the naturalness in quantum field theories. 
LHC has put strong bounds on their masses for generic superpartner spectra and decay patterns.
However, there are still search holes in the lower mass region if the SUSY spectrum is compressed. It is therefore very important to devise new methods in experimental analyses to cover these regions where the stop could still be hiding. A main difficulty in identifying the signal events is the lack of a significant MET in the stop decays if the spectrum is compressed. Recently it has been shown that this could be overcome by requiring the stop production recoiling against a hard ISR jet, which results in a large MET in the opposite direction of the ISR for the stop signal events. The experimental measurement of the $R_M(\Rmb)$ variable which corresponds to $\frac{m_{\tilde{\chi}_1^0}}{m_{\tilde{t}_1}}$ provides a powerful way to distinguish signals from the backgrounds for both all-hadronic and semileptonic decay channels. Consequently, significant regions in previous search gaps of the compressed spectrum have been excluded in the most recent ATLAS and CMS analyses.

In this paper, we extend the study to the dileptonic decay channel of the stop search. With two missing neutrinos, there are not enough kinematic constraints to solve for the $\Rmb$ to get a unique answer for a given event. However, we can find two new variables $\bar{R}_{\text{min}}$ and $\bar{R}_{\text{max}}$ which bound the interval of $\Rmb$ that is kinematically consistent with that event. In the limit of large ISR, they tend to converge to the true $\Rmb$ value. We found that these variables provide additional discriminating power, beyond the standard variables ${p_T}_{\text{ISR}}$ and MET, between the signals and backgrounds. For the traditional stop decay to $t\lsp$ in the top corridor, the dileptonic search is probably not as competitive as the all-hadronic or semileptonic channels due to the small decay branching ratios. It is not far behind though so it can still provide a complementary analysis. On the other hand, the dileptonic  search mode becomes most useful in the scenario where the stops dominantly decay through the charginos and sleptons with a compressed spectrum. In this case, the $2\ell$2$b+\slashed{p}_T$ can be the dominant final states, therefore the all-hadronic and semileptonic searches are not effective. The strong trilepton or same-sign dilepton chargino-neutralino search constraints also diminish in the compressed region. Even though the signal topology is different from the one where the kinematic constraint equations for $\Rmb$ are derived, we have shown that the $\Rmb$ variables are still useful to suppress the dominant $t\bar{t}$ background which does have the topology of the constraint equations. It can cover a significant region of the parameter space which has not been experimentally explored by other methods before.

The power of the kinematic variables and techniques studied in this paper and previous works comes from utilizing our maximal knowledge of the kinematic information of the signal and background events. It is conceivable that similar techniques can be used to extend the coverage of other compressed region where the constraint is still weak, like the $W$-corridor ($\mtt-\mlsp \approx m_W +m_b$) in the stop search, or other SUSY searches with compressed spectra. 

\section*{Acknowledgments}
We thank Zhangqier Wang for useful discussion about the tradition stop analysis.
This work is supported in part by the US Department of Energy grant DE-SC-000999. H.-C.~C. is also supported by The Ambrose Monell Foundation at the Institute for Advanced Study, Princeton, and thanks HKUST Jockey Club Institute for Advanced Study for hospitality where part of this work was done.

\appendix
\section{Solving for $\bar{R}_{\text{min}}$ and $\bar{R}_{\text{max}}$}
\label{app:technical}
In this Appendix, we explain how we determine if Eqs.~(\ref{mass_dilep}), (\ref{met_perp}) have real solutions and find the extremum values of $\Rmb$ that allow real solutions. We first guess a value for $\bar{R}_M$. Given a trial $\bar{R}_M$, we can reduce Eqs.~(\ref{mass_dilep}), (\ref{met_perp}) down to a quartic equation. To determine whether it yields real solutions, one can follow the same procedure as in an efficient method of calculating $\mathrm{M_{T2}}$~\cite{Cheng:2008hk}. We first compute the Sturm Sequence for the quartic equation and compare the number of the sign changes at the positive and negative infinities. If the number of sign changes at two infinities are different, the quartic equation yields real solutions. The point is that we can determine whether a trial $\bar{R}_M$ can solve the equations without actually solving them.

Since the $t\bar{t}$ background is more likely to have $\bar{R}_M$ close to zero, the region covered between $\bar{R}_{\text{min}}$ and $\bar{R}_{\text{max}}$ ideally should contain zero. We choose the range of the search of $\bar{R}_M$ between some negative value and some positive value. In the example of the stop-slepton decay, the search range of $\Rmb$ is chosen to be $[-5, 5]$. We first find an $\Rmb$ value which allow real solutions using the bi-section method, then look for $\bar{R}_{\text{min}}$ ($\bar{R}_{\text{max}}$) from that $\Rmb$ value by decreasing (increasing) the $\Rmb$ value with steps of $\Delta R = 0.1$.
Because the final states contain two leptons and two $b$ jets, there are two possible ways of grouping the $\ell, b$ pairs into decay products of the two top quarks. The trial point that solves at least one combination of $\ell b$ is considered a viable point. 
We scan the $\Rmb$ following the steps until reaching the point where no solutions can be found for either combination of $\ell b$. The last viable point in this scan is designated as $\bar{R}_{\text{min}}$ ($\bar{R}_{\text{max}}$).

In rare cases, there are disjoint intervals of $\Rmb$ having real solutions. We choose the $\bar{R}_{\text{min}}$ and $\bar{R}_{\text{max}}$ of the interval that contains the point 0 or is closest to 0. Specifically, if the initial $\bar{R}_{\text{min}}$ and the $\bar{R}_{\text{max}}$ determined in this way do not cover a region which contains zero, we start a new search following the same steps as described above, but with a range $(-x',x')$, where $x'$ is the smaller number between the $|\bar{R}_{\text{min}}|$ and $|\bar{R}_{\text{max}}|$ found in the last search. The search continues recursively until we reach the point where either the $\bar{R}_{\text{min}}$ and $\bar{R}_{\text{min}}$ found covers the point 0 or the range itself becomes smaller than the precision $\Delta R$.

\section{Validating the Usefulness of the $\bar{R}_M$ Variables}
\label{app:validation}

To check the usefulness of $\bar{R}_M$ as a new dimension for the dileptonic stop search, we perform a study on the stop-chargino-slepton scenario and compare the significances of an analysis mainly using ISR and MET, and an analysis including the $\Rmb$ variables. We use the benchmark BMP1 for the  numerical study and focus on the dominant $t\bar{t}$ background. 

Both signal and background events need to satisfy preliminary selection rules as mentioned in Sec.~\ref{Section:Bench}, also the leading lepton $p_T$ less than 100~GeV. For the ``control'' study without $\Rmb$ variables, we divide the ISR-MET plane into 24 non-overlapping signal regions with 4 ISR bins (200-300,300-400,400-500,500-) and 6 MET bins (200-250,250-300,300-350,350-400,400-500,500-), all in units of GeV. For each signal region the significance is calculated with a 10\% independent background systematic uncertainty. Then we compare it with an analysis which also include the variable $|\Rmbmax|+|\Rmbmin|$. The signal regions are divided into small boxes in the 3-dimensional parameter space. To have a fair comparison of the two analyses, for the 3-variable case we make a coarser grid in the ISR-MET plane, so that the total number of signal regions is also 24. Specifically, the ISR variable is divided into 2 bins (200-400,400-), MET variable is divided into 4 bins (200-250,250-350,350-500,500-) and the $|\Rmbmax|+|\Rmbmin|$ variable is divided into 3 bins (0-1,1-2.5,2.5-).

The overall combined significance for the analysis including the $|\Rmbmax|+|\Rmbmin|$ variable with $t\bar{t}$ background is $\sim 4.0$, while for the ``control'' study without $\Rmb$ variables but a finer MET/ISR binning, the overall significance is $\sim 3.1$. This corresponds to a $\sim 30\%$ improvement. 
The qualitative conclusion that the new $\Rmb$ variables help to improve the analysis holds for different choices of signal regions, whereas the extent of improvement depends on the details of signal region selections.

\end{document}